\documentclass[utf8]{frontiersSCNS} 
\usepackage{url,lineno,microtype,subcaption}
\usepackage{hyperref}
\usepackage[onehalfspacing]{setspace}
\usepackage{bm}
\usepackage[binary-units]{siunitx}

\def\keyFont{\fontsize{8}{11}\helveticabold }
\def\firstAuthorLast{M.F. Bari {et~al.}} 
\def\Authors{Md Faizul Bari\,$^{1,*}$, Parv Agrawal\,$^{1}$, Baibhab Chatterjee\,$^{1}$ and Shreyas Sen\,$^{1}$}
\def\Address{$^{1}$Department of ECE, Purdue University, West Lafayette, Indiana, USA}

\def\corrAuthor{Md Faizul Bari}

\def\corrEmail{mbari@purdue.edu}

\begin{document}
\onecolumn
\firstpage{1}

\title[Feature Selection Enhanced RF-PUF on Unmodified Commodity Devices]{Statistical Analysis Based Feature Selection Enhanced RF-PUF with \bm{$>$}99.8\% Accuracy on Unmodified Commodity Transmitters for IoT Physical Security} 

\author[\firstAuthorLast ]{\Authors}
\address{} 
\correspondance{} 

\extraAuth{}

\maketitle

\begin{abstract}
\section{}
Due to the diverse and mobile nature of the deployment environment, smart and connected commodity devices are vulnerable to various attacks which can grant unauthorized access to a rogue device in a large, connected network and lead to data theft and malicious activities. Traditional digital signature-based authentication methods are vulnerable to key recovery attacks, cross-site request forgery, etc. To circumvent the inherent weakness of the digital signature-based authentication system, RF-PUF had been proposed as a promising alternative that utilizes the inherent nonidealities of the devices as physical signatures that arise from the manufacturing process variation. RF-PUF offers a robust authentication method that is resilient to key-hacking methods due to the absence of secret key requirements and does not require any additional circuitry on the transmitter end, eliminating the need for additional power, area, and computational burden. In this work, for the first time, we analyze the effectiveness of  RF-PUF  on commodity devices, purchased off-the-shelf, without any modifications whatsoever. Data were collected from 30 Xbee S2C modules used as transmitters and released as a public dataset for the whole community. A new feature has been engineered through PCA and statistical property analysis. With a new and robust feature set, it has been shown that 95\% accuracy can be achieved using only \bm{$\sim$}1.8 milliseconds of test data fed into a lightweight neural network (with 10 neurons in 1 layer), reaching $>$99.8\% accuracy with more data and network of higher model capacity (1 layer with 70 neurons), for the first time in literature without any assisting digital preamble. The design space has been explored in detail and the effect of the wireless channel on the network performance has been determined. The performance of some popular machine learning algorithms has been tested and compared with the neural network approach in terms of device numbers. A thorough investigation on various PUF properties has been done and both intra and inter-PUF distances have been calculated. With extensive testing of 41238000 cases, the detection probability for RF-PUF for our data is found to be 0.9987, which, for the first time, experimentally establishes RF-PUF as a strong authentication method. Finally, the potential attack models and the robustness of RF-PUF against them have been discussed.

\tiny
 \keyFont{ \section{Keywords:} IoT, machine learning, physical security, PUF, neural network (NN), radio frequency, authentication, Xbee}
\end{abstract}

\section{Introduction}

The fourth industrial revolution, powered by low-power, high-speed modern communication systems has ushered in a new era of immersive and unprecedented user experience through smart devices that are connected with each other and to the cloud, popularly known as the Internet of Things (IoT). The global IoT market is experiencing a rapid boost and according to a prediction by Norton, there will be around $21$ billion connected devices by $2025$ \cite{norton}. We are already talking about the Internet of Everything (IoE) which essentially refers to people, data, and smart things connected to form an ecosystem that ensures a better and smarter lifestyle. The diverse application environment of the smart devices has rendered them vulnerable to a wide attacking surface. The weakest point in a network defines its security and for IoT networks, the resource-limited, user-end devices are the weakest points where a security compromise can provide access to a rogue device that can pose a massive threat to the whole network. So, the question of secure authentication before granting access to a large network is of increasing importance.

\begin{figure}[t]
  \centering
  \includegraphics[width=\linewidth]{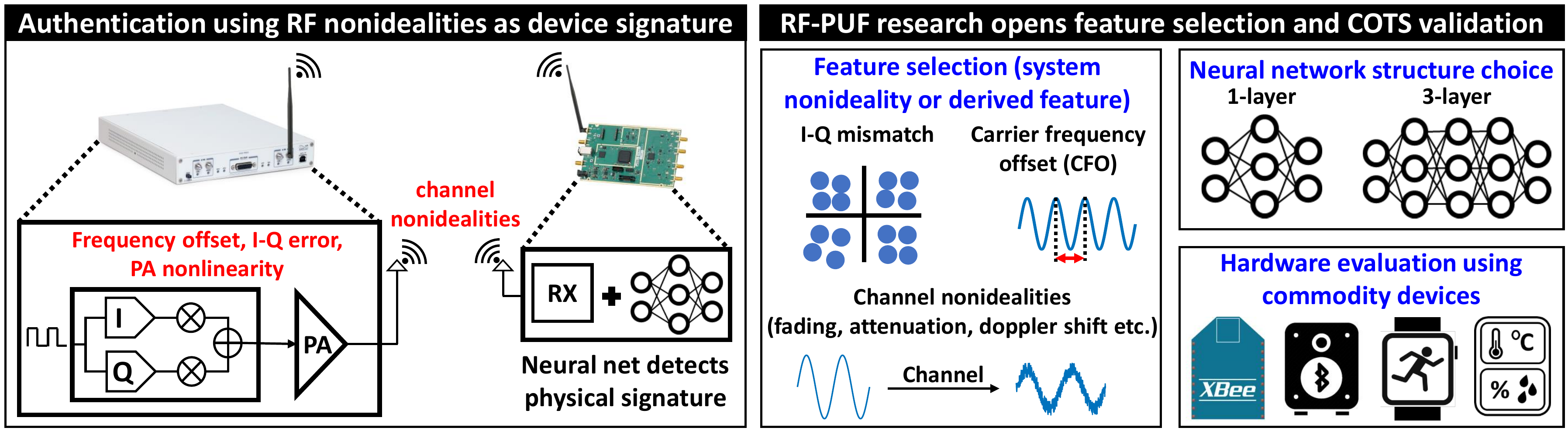}
  \caption{The concept of RF-PUF exploits the inherent physical signature embedded in the device which manifests itself as different imperfections, which are used as features to train a neural network at the receiver end for authentication. The challenges involve developing a proper feature set, choosing a proper neural network architecture, and evaluating the concept in real, commodity devices.}
  \label{concept}
\end{figure}

Traditional methods such as symmetric-key cryptography and asymmetric-key cryptography use secret private keys or public/private key pairs respectively, for encryption/decryption. Key-based methods require the storage of a secret key in a nonvolatile memory (NVM) or SRAM. However, they are vulnerable to different invasive/semi-invasive key-hacking attacks and side-channel attacks \cite{DPA, EM_SCA, ML_SCA}. Multi-factor authentication (MFA) \cite{MFA_pat, MFA_survey} requires one or more verification factors (e.g., biometric factor, two-factor code from authentication app, etc.) along with the secret key. The widely-used open authentication (OAuth 2.0) protocol \cite{oauth} for current IoT networks suffer from cross-site request forgery (CSRF) attacks \cite{csrf1, csrf2}. Both OAuth and MFA are inconvenient for large networks as they require manual verification. In addition to these vulnerabilities, the use of digital signatures also puts additional power and area burden which are typically small but could be significant for extremely energy and resource constraint edge devices.

To circumvent this, the idea of \textit{radio frequency physical unclonable function} (RF-PUF) has been recently proposed \cite{RFPUF} using physical signature instead of or in addition to the digital signature. The concept of RF-PUF is explained in Fig.~\ref{concept}. RF-PUF exploits the \textit{inherent} device imperfections due to manufacturing process variation and other system-level nonidealities (e.g. LO frequency offset, I-Q mismatch, DC offset, attenuation, fading, Doppler shift, etc.) as unique physical signatures. These signatures are used as features and fed to a neural network at the receiver to train it. Once trained, this network can be employed at the receiver for authentication. RF-PUF does not demand any additional preamble, digital keys, or assistive communication medium for authentication purposes. The absence of an external security key or preamble makes RF-PUF highly resilient to different types of key-hacking attacks and alleviates the need for preamble obfuscation \cite{PREAMBLE_OBFUS}. Also, it does not require any secured memory block for key storage. Thus, both power and area overhead is reduced on the resource-constrained edge-node side of an \textit{asymmetric} IoT network.

In \cite{RFPUF}, the idea of RF-PUF was presented primarily based on simulation data using I-Q samples as features. However, the PUF output is stochastic in nature and it is very hard to accurately capture the device nonidealities in simulation. This calls for addressing the open research needs of experimental validation of RF-PUF and demonstration of high-accuracy on devices found `in-the-wild'. In this work, we address both these research problems by a) analyzing the efficacy of RF-PUF on unmodified commodity devices and b) introducing effective feature selection to increase RF-PUF accuracy $>99.8\%$. To achieve this, an improved and robust feature set is necessary to provide a reliable authentication method. We purchased commercially available $30$ Xbee S2C devices and used them as unmodified commodity COTS (Components off-the-self) devices to experimentally validate RF-PUF. 155.4 GB of data have been collected from the Xbee transceiver systems and 2.5 GB of data have been used for experimentation. This dataset has also been made public on GitHub along with this paper, for further development and validation by the RF-Security community. 

It has been shown that $95\%$ accuracy can be achieved even with a lightweight, single-layer neural network with $10$ neurons and $\sim 1.8$ \si{\milli\second} (\SI{30}{\kilo\byte}) of test data, which ensures the feasibility of RF-PUF in a low-latency network. With statistical analysis, a new feature has been augmented that massively boosts the performance of the network. The impact of the variation in neural network model capacity and the amount of training data on detection accuracy has been explored. Along with artificial neural networks, experiments have been performed with multiple traditional machine learning algorithms, and their performance is compared in terms of the number of devices. A detailed analysis of the PUF properties has been done to evaluate the eligibility of RF-PUF as a PUF. Inter-PUF and intra-PUF hamming distances have been calculated and it has been proved that for commodity COTS (Components off the self) devices without any modification, RF-PUF shows strong identifiability with a very high ($99.87\%$) detection probability. As an authentication method, possible vulnerabilities and attack models for RF-PUF have been investigated and the robustness of RF-PUF against them has been proved. 
The insights gathered from these analyses and experiments may prove to be extremely important for the design and implementation of RF-PUF in the future in realistic application scenarios with `in-the-wild' devices.

\subsection{Our Contribution}
In this work, \textbf{through thorough statistical analysis of unmodified commodity devices we have found an optimum feature that improves the accuracy of RF-PUF significantly on a suite of commodity hardware devices leading to \bm{$>99.8\%$} accuracy, along with PUF property analysis and security vulnerability analysis.} Detailed contributions are as follows:

1) \textbf{Feature engineering:} Principal component analysis has been performed on the existing feature set found in the literature to find the dominant feature. Through moment analysis on the dominant feature (i.e. carrier frequency offset) we demonstrate that the addition of a feature called COV (ratio of standard deviation and mean of carrier frequency offset) significantly helps in achieving high ($>99.8\%$) accuracy (Section~\ref{Stat_analysis}).

2) \textbf{Highest accuracy achieved with unmodified COTS devices:} $30$ Xbee S2C modules
have been used without the help of any assisting communication preamble or any modification to the devices whatsoever. Using data received over a wireless channel with a suitable feature set and a lightweight neural network, $99.8\%$ accuracy can be achieved which, to our best knowledge, \textit{is the highest accuracy using this many commodity COTS devices considering the wireless channel} (Section ~\ref{COV_result}).

3) \textbf{RF-PUF established as a strong PUF:} Any distinct PUF class is identified through some properties that make it a separate class. They include constructability, evaluability, uniqueness, reliability, and identifiability. We have explored these properties for RF-PUF in detail, calculated intra-PUF and inter-PUF hamming distances and in an extensive test of $41238000$ cases, we have shown that the probability of proper identification of an RF-PUF instance is $0.9987$. This is the first time analysis of RF-PUF as a PUF class which experimentally demonstrates RF-PUF as a strong and unique PUF class by itself (Section~\ref{PUF_properties}).

4) \textbf{Performance evaluation using popular machine learning algorithms} and comparison with neural network (NN) based approach. It has been shown that even a lightweight NN with a single hidden layer can handle $>$300 devices with 99.9\% accuracy, unlike ML algorithms (Section~\ref{ML_and_NN}).

5) \textbf{Wireless channel variability analysis} on the accuracy of RF-PUF and the effect of network depth on accuracy with and without wireless channel has been presented. Discussion on possible important attack models and the robustness of RF-PUF against such attacks (Section~\ref{ch_effect}).

6) \textbf{Public Dataset:} Our collected data have been released as a public dataset for the whole community to explore and experiment with (Section~\ref{dataset}).

\begin{figure}[t]
  \centering
  \includegraphics[width=\linewidth]{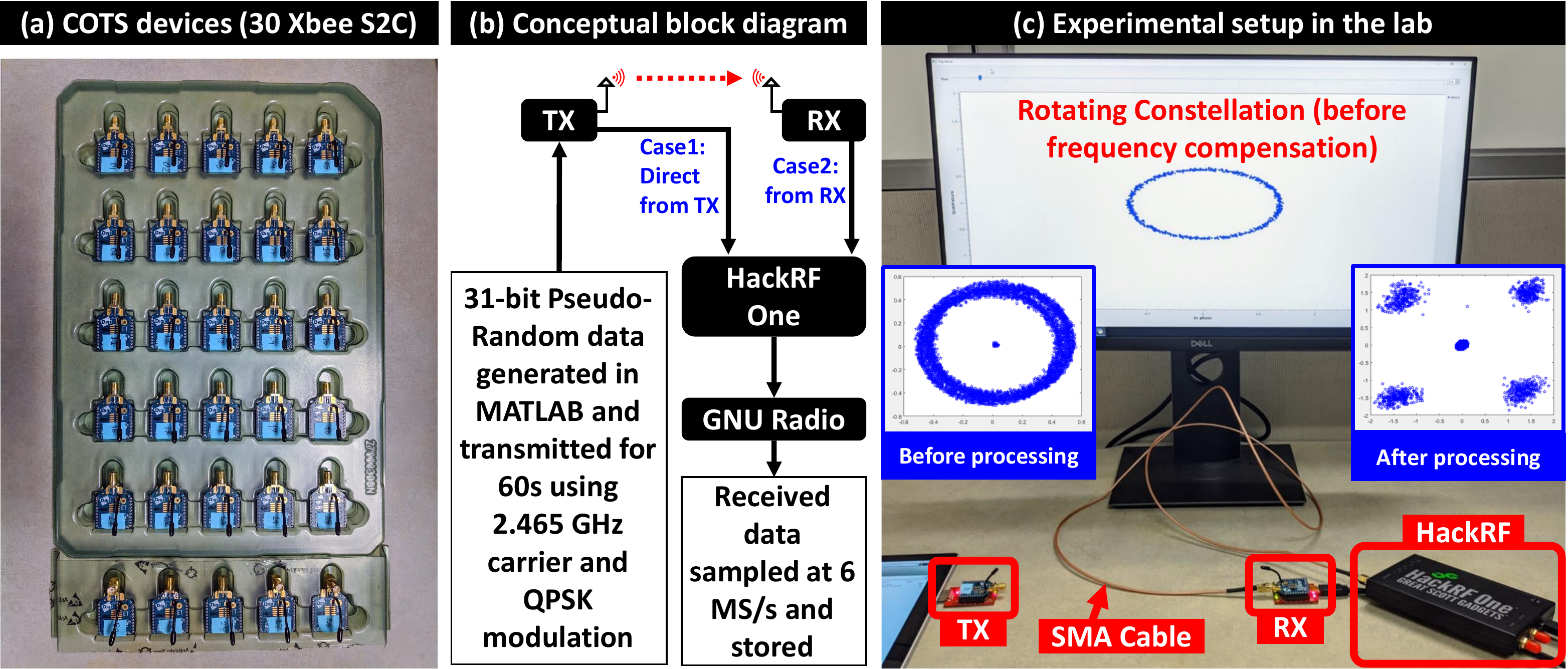}
  \caption{(a) Commodity off-the-shelf devices (30 Xbee S2C modules) used for data collection. (b) Conceptual experimental setup. (c) Actual experimental setup in the lab. The TX and RX are placed \SI{1}{\meter} apart (they are close here for image capturing purpose only) and a HackRF module was used to collect data either from the TX (case 1) or RX (case 2). GNU Radio records the collected data and shows a live constellation (visible on-screen). The rotating constellation is later processed in MATLAB through coarse and fine frequency compensation.}
  \label{setup}
\end{figure}
\begin{figure}[t]
  \centering
  \includegraphics[width=0.65\linewidth]{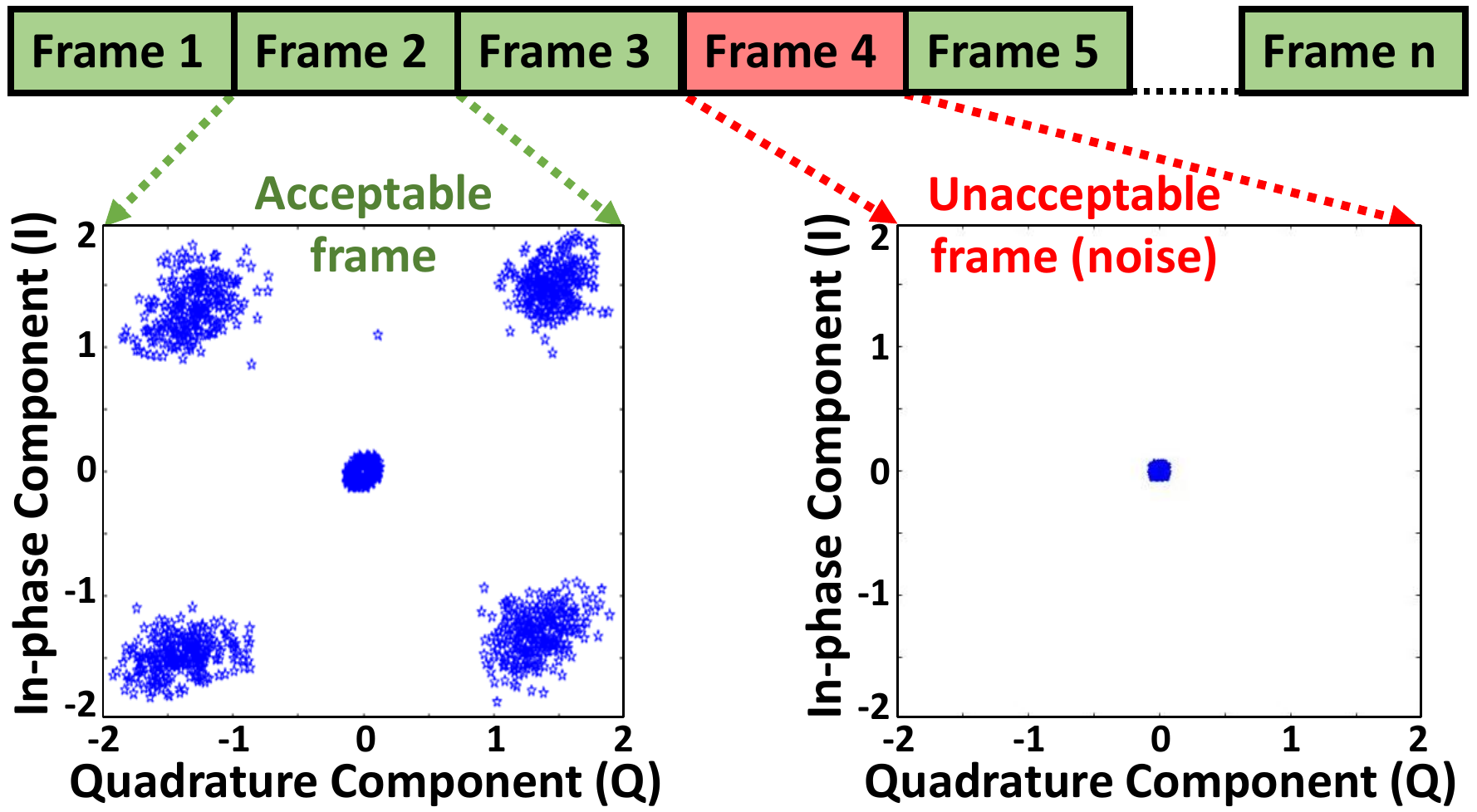}
  \caption{Grouping collected data in a number of frames and filtering of data for acceptable frames. This step is required as the Xbee module transmits data on an interval.}
  \label{filterframe}
\end{figure}
\section{Related Works}
Time and frequency domain properties of individual transmitters have been used for RF fingerprinting \cite{RFfingerBT, RFfingerFC, RFfingerXbee, RFfinger_4, RFfinger_5, RFfinger_6, RFfinger_7, RFfinger_8, RFfinger_9}. However, both time and frequency domain analysis have their limitations in the form of detecting the start and end of the transients, high oversampling ratios, and the need for fixed preambles to avoid data dependency. MAC layer and other upper layers of the communication protocol have also been used for RF-fingerprinting \cite{RFfingerMACLayer}. However, device identifiers in upper layers like IMEI number, IP address, MAC address, etc. can be spoofed \cite{IMEI, IP_hack, http_hack, MAC_spoof}.
Hanna et al. utilized power amplifier nonlinearity to fingerprint RF devices \cite{PA_nonlinear} using simulation data. Recently, there have been a growing number of works that use raw RF data and depend on complex neural networks to classify devices \cite{DL_1, DL_3, DL_4, DL_5}. This method has one weak point. As wireless data are contaminated with noise and interference, any use of the RF data without processing always posits a risk of huge performance drop in scenarios where environmental nonidealities can go beyond the estimation that was used while designing the network. Contaminated data can render faulty predictions, especially if the training environment is significantly different from the test environment. Another concern in this approach is that it is somewhat blind and does not provide intuition on different design parameters and their effects. Processing data, extracting a proper feature set, and unraveling the mystery of the design space can render a robust authentication method that is less vulnerable to environmental factors and provides more flexibility to the designer. That's why RF-PUF performs better than the CNN-based approach as is shown in \cite{IMS}.
Brik et al. \cite{IEEE_802} used IEEE 802.11 devices to show $99\%$ detection accuracy. However, the wireless channel was ignored completely. In our work, using data from $30$ commercial Xbee devices and considering the wireless channel in conjunction with a lightweight NN, we have shown that we can achieve $>99.8\%$ accuracy.

\section{Experimental Setup}
\subsection{Physical Device Setup}
For experimental validation, 30 XBee S2C modules are chosen (IEEE 802.15.4 standard) which is designed for industrial and commercial use. Fig.~\ref{setup}(a) shows the Xbee devices whereas Fig.~\ref{setup}(b), and Fig.~\ref{setup}(c) show the block diagram and the actual setup. The TX and RX are kept 1 m apart. Using SMA cable, a HackRF One software-defined radio (SDR) module has been connected either to the TX (case 1) or to the RX (case 2) to extract data excluding (case 1) or including (case 2) wireless channel.

\subsection{Data Collection and Filtering Noise}
A 31-bit pseudo-random bit sequence (PRBS) is generated in MATLAB and fed to each TX which transmits this data for 60 sec with QPSK modulation at 2.465 GHz and 230400 bps baud rate. These data were captured in a Xbee RX module. Simultaneously, data were also captured by a HackRF one software-defined radio (SDR) module, sampled at 6 MSps, and stored by GNU Radio. The captured data are divided into several frames, each containing a number of samples.
From the constellation diagram of the frame data (Fig.~\ref{filterframe}), it is found that some frames have no significant data points and contain only noise as the Xbee devices transmit data intermittently due to its buffer limitation. These blank frames containing only noise were discarded.

\subsection{Public Dataset}
\label{dataset}
This dataset contains raw data collected from 30 Xbee S2C transmitters for both cases (excluding and including the channel) in binary format. The total size of the dataset is 155.4 GB (each transmitter data is $\sim$2.5 GB). It can be downloaded from `Sparclab RF-PUF Dataset \cite{RF_PUF_Dataset}'.

\begin{figure}[t]
  \centering
  \includegraphics[width=\linewidth]{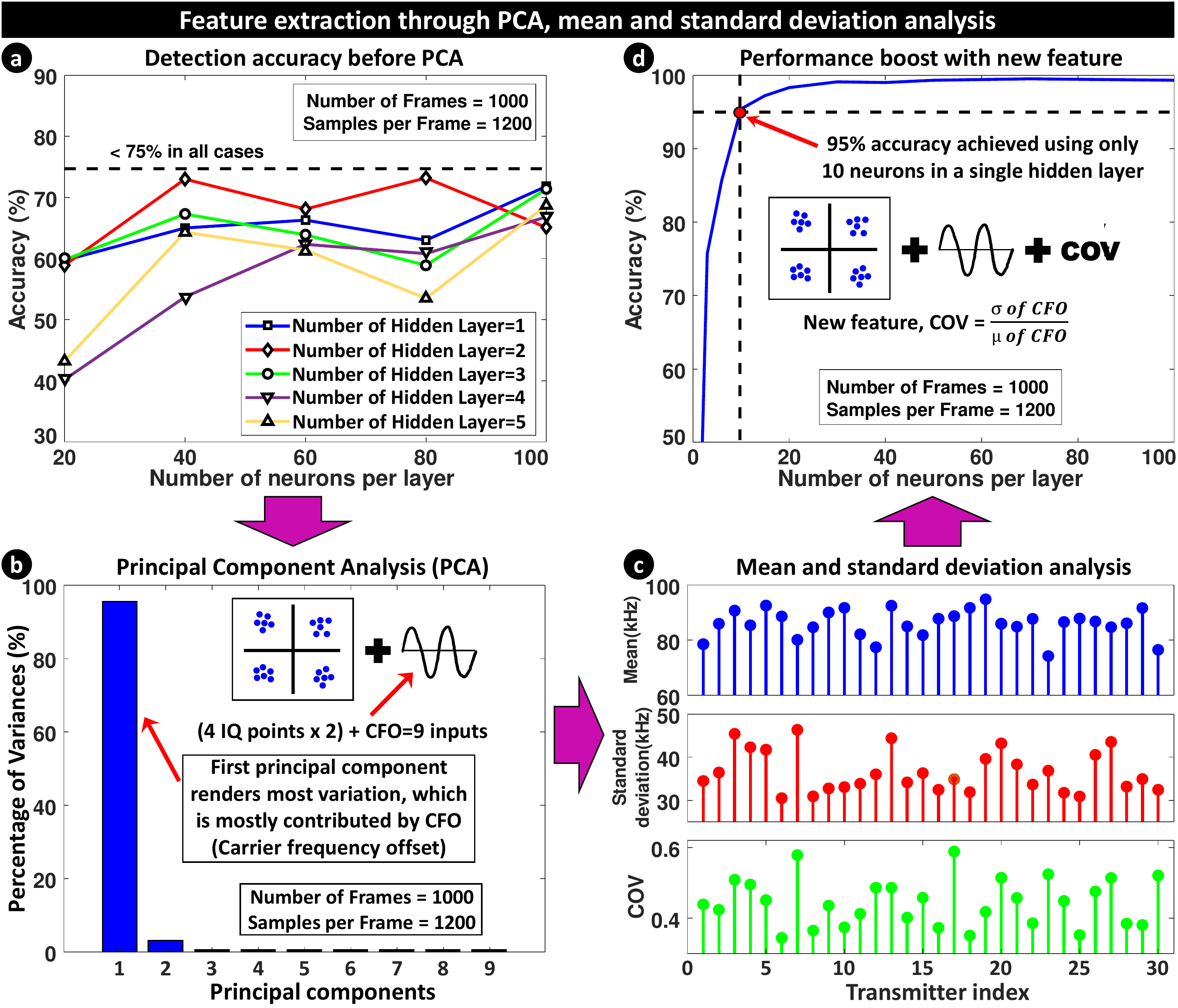}
  \caption{(a) Accuracy vs the number of neurons in each hidden layer. Even after increasing the number of hidden layers, the accuracy remains $<75\%$. (b) Principal Component Analysis (PCA) reveals that first principal component (PC) causes most variation, which in turn depends mostly on the carrier frequency offset, CFO. (c) Mean ($\mu$) and standard deviation ($\sigma$) of the dominant feature (CFO) analyzed in search of a new feature. It reveals that these statistical parameters vary significantly among transmitters. So, their ratio or coefficient of frequency offset variation, COV = $\frac{\text{standard deviation ($\sigma$) of CFO}}{\text{mean ($\mu$) of CFO}}$ is taken as the tenth feature. (d) Inclusion of COV shows significant improvement in the detection accuracy. Using a single hidden layer with only $10$ neurons, $95\%$ accuracy is achieved, and $>99\%$ accuracy is reached for $>50$ neurons.}
  \label{init_acc}
\end{figure}

\section{Feature Extraction}
\subsection{Initial Feature Set}
In our work, CFO and I-Q data are taken as features just as in the original RF-PUF paper\cite{RFPUF}. The previously generated frames are filtered using matched filtering, frequency compensated (both fine and coarse), and finally synchronized using timing recovery. In this process, CFO is found as a byproduct. Along with CFO, the compensated in-phase and quadrature-phase components in four quadrants are used as features. The $9$ features (CFO $+$ 4 I-components $+$ 4 Q-components) from each frame and $1000$ frames from each TX lead to a feature set of $9\times 1000$. The final feature matrix is a combination of these feature sets from all $30$ devices and has a size of $9\times 30000$.

\subsection{Accuracy with Carrier Frequency Offset and I-Q Features}
The whole feature data are divided into $70\%$, $15\%$, and $15\%$ respectively for training, validation, and test purposes and fed into a neural network (NN). The performance of the neural network is tested by varying the number of neurons and hidden layers. Fig.~\ref{init_acc}(a) shows the accuracy of the trained model for different neural networks. The accuracy is less than $75\%$ in all test cases. Since exploring different NN configurations does not provide expected accuracy, our choice here is to: (i) form an improved feature set to be used with the NN (ii) use different machine learning (ML) algorithms (iii) use more data. We first search for an improved feature set for better accuracy. Later, effect of more data is shown in subsection~\ref{more_data1},~\ref{more_data2} and a comparison of different ML algorithm and NN is discussed in subsection~\ref{ML_and_NN}.

\subsection{Statistical Analysis}
\label{Stat_analysis}
\subsubsection{Principal Component Analysis}
We start the investigation by performing Principal Component Analysis (PCA) with feature matrix as input (each feature represents one input dimension). Fig.~\ref{init_acc}(b) shows the principal components and their contribution to the variances. The first principal component (PC) contributes to most of the variances and the input to PC mapping reveals that the CFO is the most dominant feature. So, an in-depth statistical property analysis of the CFO can help in deriving a new feature.

\subsubsection{Moment Analysis} Since, CFO varies from frame to frame (i.e., with time), it is intuitive to look at the moments of their distribution. Specifically, we want to look at first and second-order moments (mean and variance). Fig.~\ref{init_acc}(c) shows the absolute values of mean and standard deviation (square root of variance) of CFO. These parameters vary significantly from TX to TX in most cases. And even if for any two TX, the mean is similar, the standard deviation is different, and vice versa. If they can be combined to form a new feature, that can provide significant discrimination among transmitters and lead to much better accuracy. In statistics, the ratio of standard deviation and mean is known as the coefficient of variation. So, using this statistical parameter, we form a new feature named the coefficient of frequency offset variation (COV) which is defined as:
\begin{equation*}
    COV = \left|\frac{\textit{Standard deviation of CFO}}{\textit{Mean of CFO}}\right|
\end{equation*}

\subsection{Performance using COV feature}
\label{COV_result}
COV is included as the tenth feature in our existing feature matrix. From PCA analysis, it is already revealed that the I-Q features contribute to much fewer variances and can be discarded by trading some accuracy. Since our goal is to achieve maximum possible accuracy, we still keep them as features. Also, I-Q values contain channel information, which will help the NN to compensate for the wireless channel (channel effect is explained in subsection~\ref{ch_effect}).

After including COV as the tenth feature, our neural network was trained, validated, and tested again with the new feature matrix. Fig.~\ref{init_acc}(d) shows that the performance of the network has improved drastically. With just a single hidden layer, $>95\%$ accuracy can be achieved using $10$ neurons and can hit $99.9\%$ accuracy by increasing the number of neurons.

\section{Evaluation of Design Parameters}
\begin{figure}[t]
  \centering
  \includegraphics[width=\linewidth]{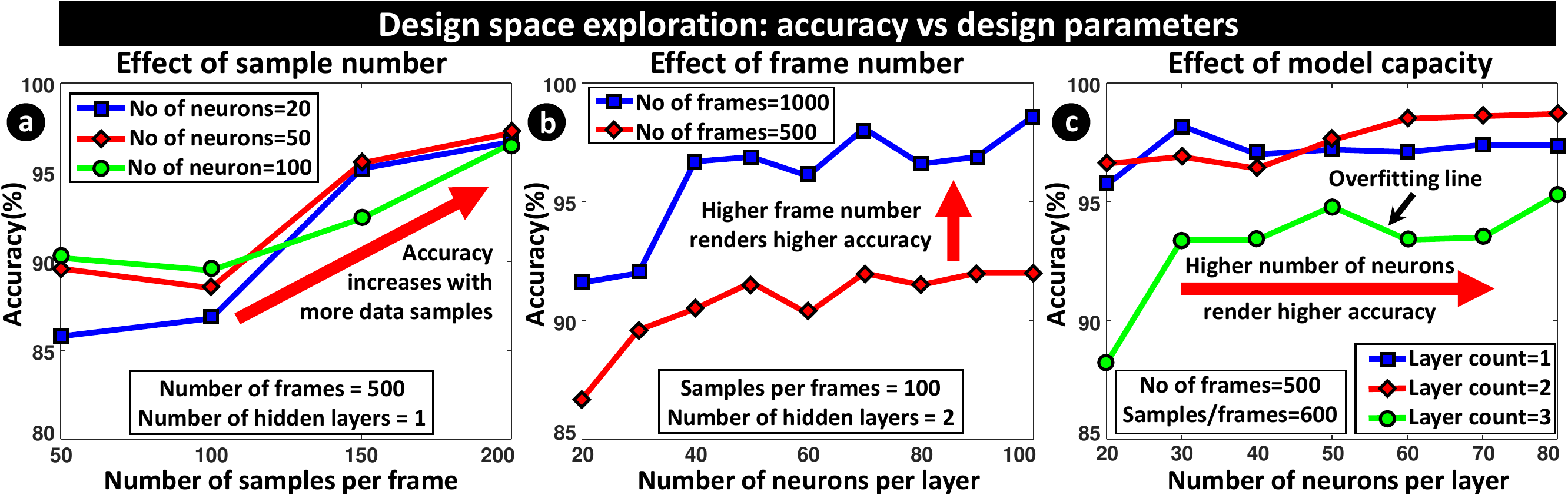}
  \caption{(a) Detection accuracy vs the number of samples per frame for different neural networks which shows a trend of accuracy improvement (indicated by red arrow) with the increase in sample number. (b) Detection accuracy for different frame numbers shows higher frame number renders better performance. (c) Detection accuracy versus the number of neurons per layer. The general trend shows that accuracy improves with the increase in the number of neurons in each layer. Also, accuracy typically improves with more hidden layers until the higher model order causes overfitting and degrades performance.}
  \label{design_space}
\end{figure}

\subsection{Effect of number of samples}
\label{more_data1}
Fig.~\ref{design_space}(a) shows the plot of detection accuracy versus the number of samples in each frame for different neural networks. The general trend is that, for each NN configuration, detection accuracy improves with the increase in the number of samples (along the x-axis). This is expected because a higher number of samples provide more information and hence better performance. Also, $>95\%$ accuracy point is reached at around $150$ samples per frame which is equivalent to \SI{12.5}{\milli\second} of total data (or \SI{1.8}{\milli\second} test data). Hence, we can reach the $95\%$ accuracy bar using small test data.

\subsection{Effect of the Number of Frames in Feature Set}
\label{more_data2}
Fig.~\ref{design_space}(b) shows accuracy for two different frame numbers, $500$ and $1000$. With a higher frame number, the information content of each transmitter device increases and so their detection gets better as shown by the blue ($1000$ frames) and brown lines ($500$ frames) respectively. We can generalize the previous subsection (sample number effect) and this subsection as more data render better performance.

\subsection{Effect of the Neural Network Parameters}
Fig.~\ref{design_space}(c) shows the plots of accuracy versus the number of neurons in each hidden layer. As the number of neurons increases along the x-axis, accuracy, in general, gets better. Also, as the number of hidden layers increases, the network performs better initially, but later it creates an overfitting problem where the model capacity is too large compared to data. This phenomenon directly manifests itself as a degradation in performance. Hence, there is an optimum model capacity up to which accuracy increases, and beyond that accuracy drops.

\begin{figure}[t]
  \centering
  \includegraphics[width=0.7\linewidth]{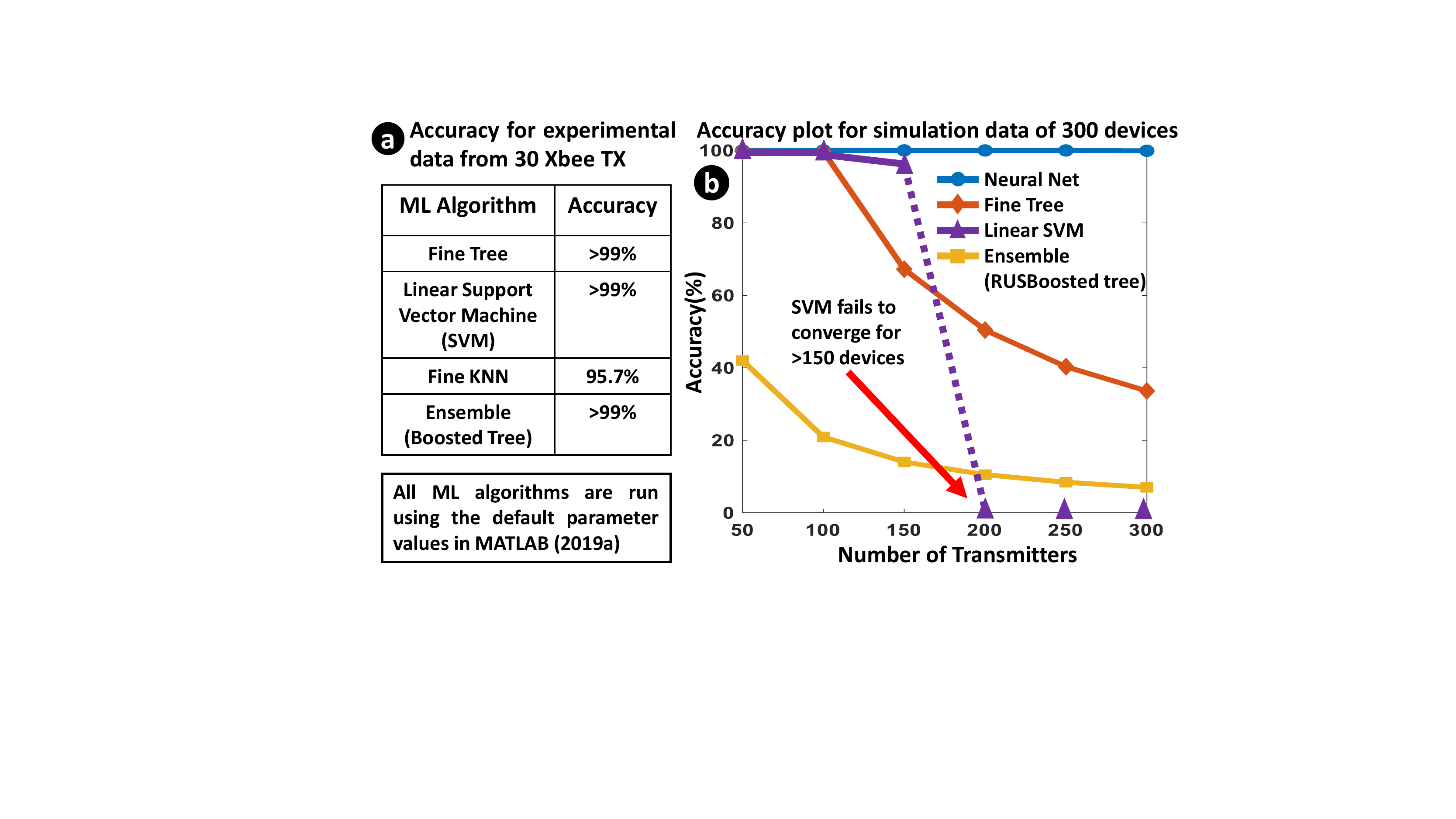}
  \caption{(a) Popular ML algorithms show high accuracy for 30 Xbee devices. (b) The accuracy of simple ML networks drops when the number of TX is large, wherein the neural network still holds up with $>99.9\%$ accuracy.}
  \label{ml_vs_nn}
\end{figure}

\subsection{Using Simple Machine Learning (ML) Algorithm}
\label{ML_and_NN}
It has been observed that the COV values vary significantly among different transmitters. When a simple feature displays a significant separation among different classes, it can be modeled with a complex \textit{if-else ladder} structure. This implies that even simple ML algorithms (e.g. Tree) can show good results. Fig.~\ref{ml_vs_nn}(a) shows that some popular ML algorithm achieves $>95\%$ accuracy.

The true power of the neural network comes into play when the number of TX increases as shown in Fig.~\ref{ml_vs_nn}(b). For this, features are generated for $300$ TX devices following a Gaussian distribution (as in \cite{RFPUF}) with the same mean and variance as that of the original $30$ TX devices, for both inter and intra-class variations. Fig.~\ref{ml_vs_nn}(b) shows that as the number of TX increases, accuracy falls after a certain point ($\sim100$ TX) even for support vector machines (SVM), and it fails to converge for $>150$ TX. 

\subsection{Effect of Wireless Channel}
\label{ch_effect}
\begin{figure}[h]
  \centering
  \includegraphics[width=0.55\linewidth]{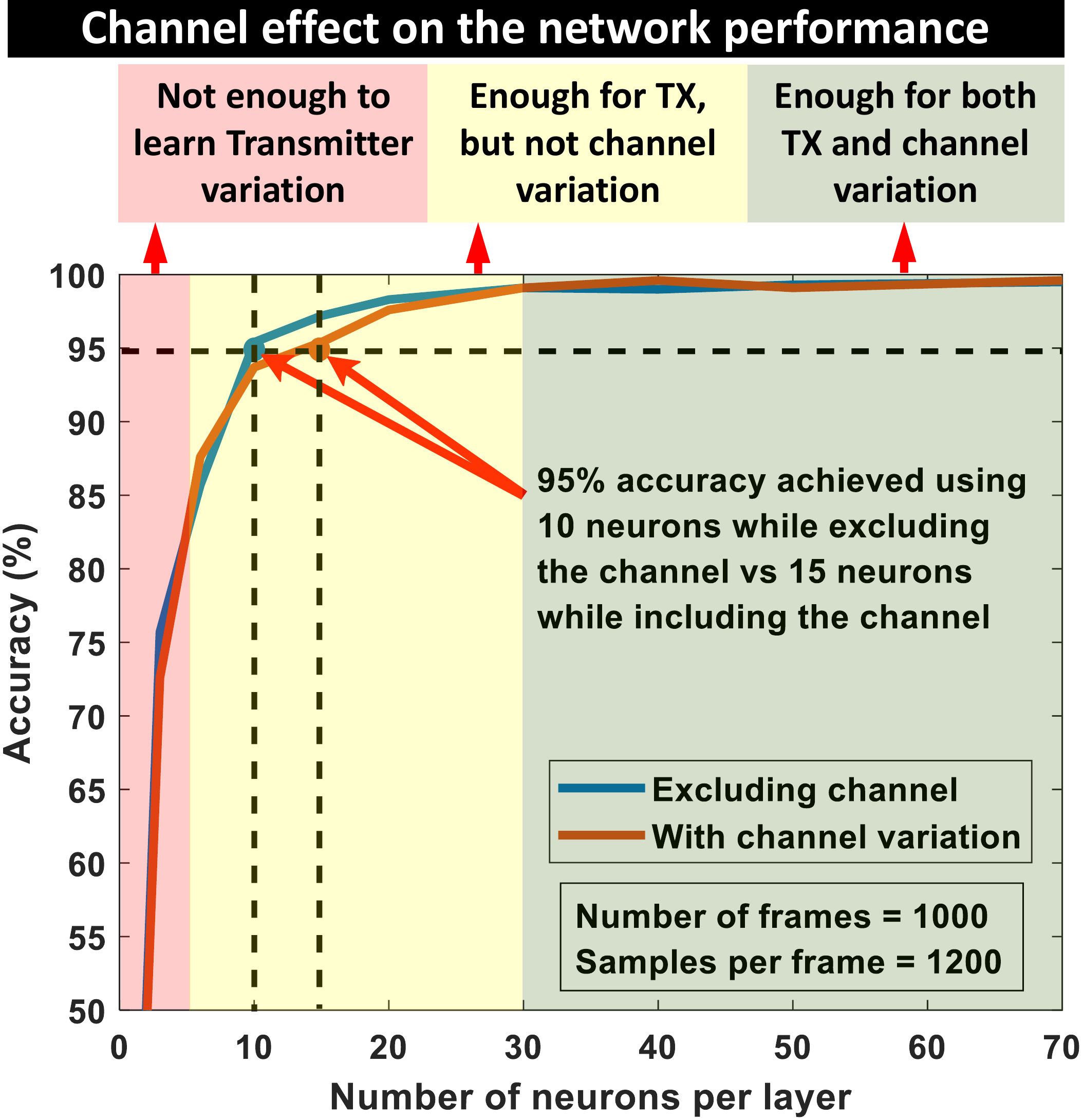}
  \caption{Comparison of the network performance in the cases of including and excluding the wireless channel data. The network needs 15 neurons compared to 10 neurons in a hidden layer to achieve $95\%$ accuracy for the case where the channel is considered. But with higher model capacity, both lines converge and the network learns the channel effect on data. d the network learns the channel effect on data. The light red box shows the region where the network fails to learn transmitter variation, light yellow box shows the region where the network learns transmitter variation but fails to learn the variation due to the wireless channel. The light green box shows the region where the network learns both the transmitter and channel variation properly.}
  \label{acc_with_ch}
\end{figure}

So far, nonidealities due to TX were considered and the wireless channel was ignored (TX, RX connected via SMA cable). But the channel itself adds some nonidealities. Here, the effect of a static wireless channel (\SI{1}{\meter} of fixed TX-RX separation) has been analyzed. Fig.~\ref{acc_with_ch} shows accuracy versus neuron number in a single layer, with and without the wireless channel. For iso-accuracy of $95\%$, wireless channel demands slightly higher model capacity ($10$ vs $15$ neurons). But when the number of neurons increases ($>50$), both curves merge and render similar accuracy.

In one of our recent works \cite{IMS}, we applied RF-PUF on the ORACLE dataset which contains data for 16 USRP X310 TX for both static and quasi-static (variable TX-RX separation) cases with a channel length varying from 2ft to 62ft. We have shown that RF-PUF achieves $100\%$ accuracy up to 38ft and $>95\%$ accuracy even at 62ft channel length. This result confirms that the RF-PUF approach can make the channel compensation with the help of NN and render high performance even in a long wireless channel. On a side note, that work combined with current work, also confirms that RF-PUF achieves high accuracy on experimental data in different platforms (XBee vs USRP radios using WiFi) for different devices.

\section{Analysis of PUF Properties}
\label{PUF_properties}
PUF response to a particular challenge is a probabilistic function. In this section, we will determine intra-PUF hamming distance and inter-PUF hamming distance and discuss various PUF properties (\cite{roel2012physically, HOST_course}) in light of those distances.

\subsection{Constructability}
A PUF class $\mathbb{P}$ is constructible if we can create a new PUF instance $puf_m \in \mathbb{P}$ through a process, $\mathbb{P}$.Create : $puf_m \leftarrow \mathbb{P}\text{.Create}$, where $puf_m$ has entropy that makes it distinct from other PUF instances $puf_{n,n \ne m}$.
In the case of RF-PUF, the source of entropy is the manufacturing process variation. During fabrication of ICs, we have within die and die-to-die variation which is due to the limitation of the manufacturing process. In contrast to many other PUF classes where we need a separate mechanism for PUF instance creation, the manufacturing process of the integrated circuit itself serves as the creation process for RF-PUF which is one of its advantages.

\subsection{Evaluability}
A PUF class $\mathbb{P}$ is evaluable if for a random PUF instance $puf_m \in \mathbb{P}$ and a random challenge $(x)$, we can evaluate a response $y:y \leftarrow puf_m(x)$. In our case, the challenge is a randomly generated bitstream in MATLAB that is fed into the transmitter and the corresponding response is the analog signal that contains the unique physical signature of the transmitter. 

\subsection{Inter PUF Distance - Uniqueness}
Uniqueness refers to how different each instance of a PUF class $\mathbb{P}$ is from each other. A measurement metric that is used to represent PUF uniqueness is called inter-PUF hamming distance and is defined as:
\begin{equation*}
 HD_{inter} \cong distance[Y_m^\alpha(x), Y_n^\alpha(x)]
\end{equation*}
 
Here, $Y_m^\alpha(x)$ and $Y_n^\alpha(x)$ are the responses from $puf_m$ and $puf_n$ (two instances of PUF class $\mathbb{P}$) under same environmental condition $\alpha$ and same challenge x. Ideally, these inter-chip hamming distances should be much greater than any intra-chip hamming distances to distinguish them separately.
In our experiment, our PUF class $\mathbb{P}$ = RF-PUF and $puf_i, (\text{where } i=1,2,...30)$ are the instances of that class (30 Xbee devices). 

To calculate $HD_{inter}$, the first $1000$ frames from each of the transmitters are taken. Each frame contains $3600$ samples. Our features remain unchanged: CFO, eight I-Q component values, and COV. But after taking $10$ features from each of $1000$ frames, instead of using them as a feature matrix for each transmitter, the mean values of the features are taken across all the frames. This means that instead of representing each transmitter as a $10\times1000$ feature matrix, it is represented as a $10\times1$ feature vector. The reason for taking the average value across the frames is that the frames have an associated timestamp with them i.e., each frame data are collected from time to time. So, each frame faces slightly different environmental conditions such as heating of the transmitter due to data transmission for a long time, external interference, noise, etc. Averaging the feature values across a large number of frames mitigates the environmental factors, especially noise. Also, taking the first $1000$ frames from each transmitter ensures the same initial heating pattern across devices. So the final outcome is that the feature vector for each transmitter has a very similar environmental factor $\alpha$, which is one of the conditions of inter-chip hamming distance calculation.

After taking feature vector from each transmitter, the Euclidean distance was calculated in ten dimensional feature space as hamming distance. For $puf_m$, let us denote $CFO_m$ = carrier frequency offset, $COV_m$ = coefficient of frequency offset variation, $I_{k,m}$ = in-phase component in the $k-th$ quadrant and $Q_{k,m}$ = quadrature-phase component in the $k-th$ quadrant. Then distance $d_{m,n}$ between $puf_m$ and $puf_n$ instances is given by:

\begin{equation}
\begin{split}
    d_{m,n}^2 = & (CFO_m-CFO_n)^2 +
    (COV_m-COV_n)^2 +\\ & \sum_{k=1}^{4}(I_{k,m}-I_{k,n})^2 + \sum_{k=1}^{4}(Q_{k,m}-Q_{k,n})^2
\end{split}
\label{norm}
\end{equation}

The inter-chip distances were calculated for each transmitter with respect to all $30$ transmitters (including the chip under test), which leads to a $30\times30$ symmetric matrix (upper and lower triangular matrices with same values since $d_{m,n} = d_{n,m}$ = inter-chip distance between $puf_m$ and $puf_n$) with a principal diagonal of zeros (self-distance). It is found that the worst case scenario with minimum distance, $HD_{inter,min}=0.2307$ and the best case scenario with maximum distance, $HD_{inter,max}=10.149$.

\begin{figure*}[h]
\centerline{\includegraphics[width=\linewidth]{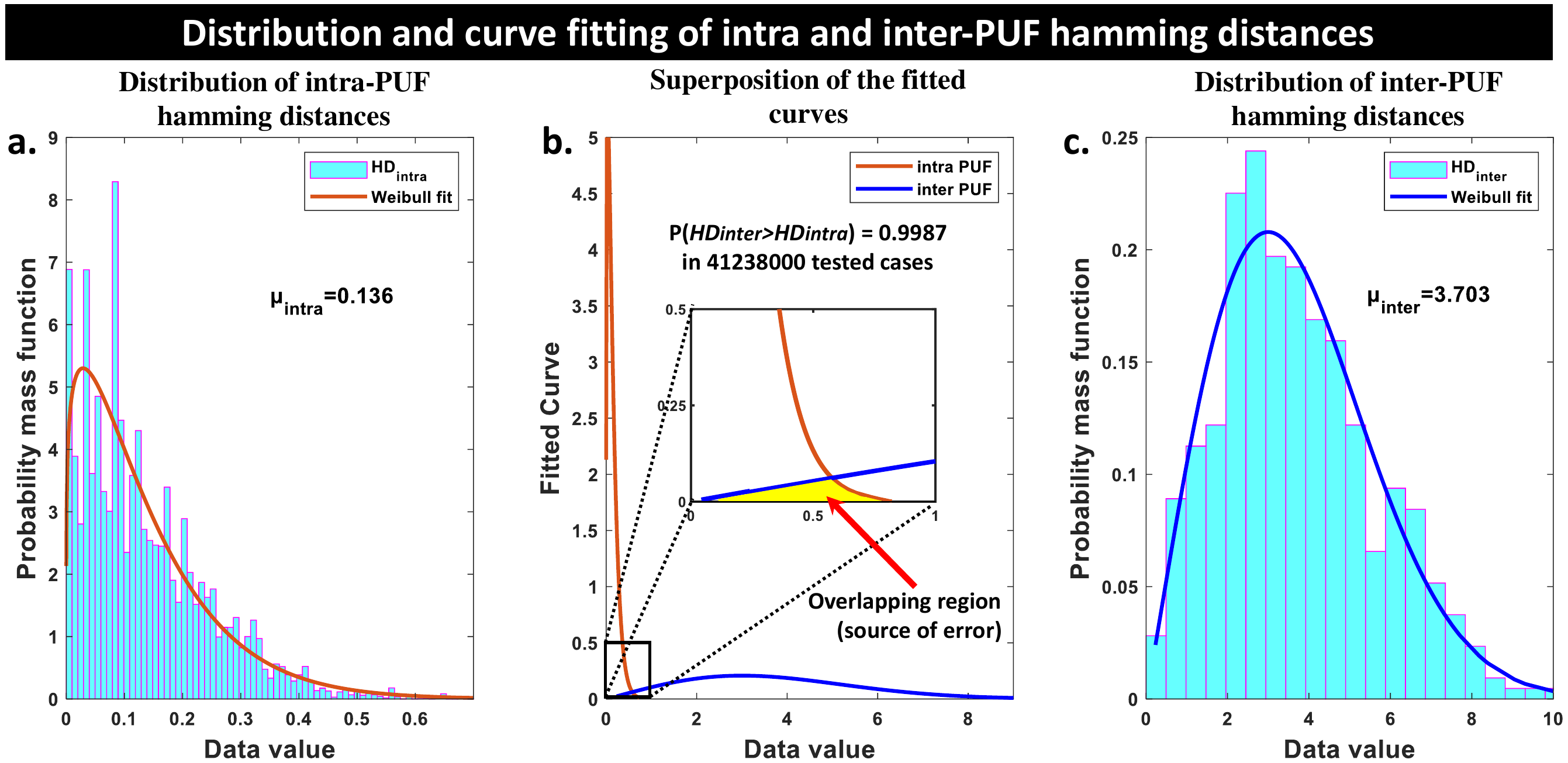}}
\caption{Data distribution of (a) intra-PUF hamming distances and (c) inter-PUF hamming distances. Due to skewness, Weibull distribution fitting is a more accurate representation in these cases. (b) The two Weibull curves are superimposed on top of each other. It is seen that there is a very slight overlap (yellow region) between the curves which is shown in a zoomed inset. Although trivial, this overlapping is the source of the detection error.}
\label{distribution}
\end{figure*}

In literature, often a mean inter-puf distance, $\mu_{inter}$, is reported which is the average of all $HD_{inter}$. The formula is:
\begin{equation*}
\begin{split}
    \mu_{inter} & = \overline{HD_{inter}} \\
    & = \frac{2}{N_{puf}\times(N_{puf}-1)\times N_{chal}}\sum HD_{inter}
\end{split}
\end{equation*}
Where $N_{puf}$ is number of puf instances ($N_{puf}=30$ for us), and $N_{chal}$ is the number of challenges ($N_{chal}=1$, since we are not varying our challenge). Using this formula, we find that $\mu_{inter}=3.703$.

Fig.~\ref{distribution}(c) shows the probability mass function distribution of $435\text{ } (=\frac{30\times29}{2})$ inter-PUF distances. The density function is right-skewed, that's why Weibull fitting (which is exponential in nature) fits it more accurately than normal distribution fitting. This fitting shows that on the right side the curve is more sparse but on the left side it is more centered instead of being sparse, which is good because that will ensure that the inter-PUF values don't go to overlap intra-PUF distances which should ideally be at zero. 

\subsection{Intra PUF Distance - Reliability}
PUF responses are in general dependent on various environmental factors that render any PUF instance response as a probabilistic function. This means that a particular PUF instance can provide slightly different values of features based on varying environmental conditions. For authentication purposes, this poses an issue. Reliability refers to how resilient a PUF instance is against environmental factors e.g. noise, interference, temperature, supply voltage, etc.

A measurement metric that is used to represent how reliable a particular instance of a PUF class $\mathbb{P}$ is intra-puf hamming distance and is defined as:
\begin{equation*}
 HD_{intra} \cong distance[Y_m^\alpha(x), Y_m^\beta(x)]
\end{equation*}
Here, $Y_m^\alpha(x)$ and $Y_m^\beta(x)$ are the responses from $puf_m$ under two distinct environmental conditions $\alpha$ and $\beta$ and same challenge x. Many $HD_{intra}$ distances are calculated at different environmental conditions. Ideally this intra-chip hamming distances should be zero.

To calculate $HD_{intra}$, we follow two steps. Let us consider one particular PUF instance $puf_m$. On step 1, the first $1000$ frames (frame number $1$ to frame number $1000$) were taken from $puf_m$, each frame containing $3600$ samples. Then mean values of the previously mentioned ten features were taken just as before to represent it as a $10\times1$ feature vector. Let us represent this vector as $f_{v,1}$. Then on step 2, first $5$ frames are skipped and the next $1000$ frames are taken from frame number = $6$ to frame number = $1005$. Step 1 is repeated here to get next feature vector $f_{v,2}$. Then next $1000$ frames are taken from frame number = $11$ to frame number = $1010$ and a feature vector $f_{v,3}$ is formed. This process is repeated $80$ times to form $80$ different feature vectors $f_{v,\alpha};\text{ } \alpha=1,2,...,80$. These $10\times1$ feature vectors are stacked together to form a feature vector set $f_{set,m}$ of size $10\times80$ for $puf_m$. The whole process is then repeated for all $30$ devices.

The purpose of taking frame-shifted or time-shifted frame groups is to consider the time factor. Each frame has a duration of $0.6ms$, so $5$ frames gap in between two frame groups renders a time difference of at least $3ms$ (in reality the difference is much larger since the transmitter transmits data for a small time and most of the frames are just noise which are filtered in data pre-processing step). The $80$ time-spaced frames, in reality, cover almost half a minute. Our $2.4GHz$ clock will have LO drift cycle time in the nanoseconds range. Hence, half-minute data can incorporate significant environmental factors into frame data. So, it can be assumed that the feature vectors $f_{v,\alpha};\text{ } \alpha=1,2,...,80$ in feature vector set $f_{set,m}$ of $puf_m$ represents $\alpha=80$ different environmental conditions.

Now, for each instance $puf_m$, Euclidean distance is calculated in $10$ dimensional feature space among the feature vectors in the feature vector set using Eq.~\ref{norm}. This results in a symmetric matrix of size $80\times80$ with a principal diagonal of zeros. This process is repeated for other transmitters as well. Essentially it gives us $30$ matrices of size $80\times80$ for intra-PUF distances. In the best-case scenario, the minimum distance is $HD_{intra,min}=7.23\times10^{-5}$ and in the worst case scenario, the maximum distance is $HD_{intra,max}=0.73$. 

Fig.~\ref{distribution}(a) shows the probability mass function distribution of $94800\text{ } (=\frac{30\times80\times79}{2})$ intra-PUF distances. The density function is right-skewed and Weibull distribution gives better fitting for it just like inter-PUF cases. This fitting shows that on the left side the curve is strongly centered towards zero, but has a diminishing trail on the right. this trail goes on to overlap inter-puf distances slightly and causes a few detection errors. Detection probability is discussed in the next subsection.

Finally, a mean intra-PUF distance, $\mu_{intra}$, is calculated which is the average of all $HD_{intra}$. The formula is:
\begin{equation*}
\begin{split}
    \mu_{intra} & = \overline{HD_{intra}} \\
    & = \frac{2}{N_{puf}\times N_{chal}\times \alpha \times (\alpha-1)}\sum HD_{intra}
\end{split}
\end{equation*}
Where $N_{puf}$ is number of puf instances ($N_{puf}=30$ for us), $N_{chal}$ is the number of challenges ($N_{chal}=1$, since we are not varying our challenge) and $\alpha$ is number of environmental conditions ($\alpha=80$ in our study). Using this formula, it is found that $\mu_{intra}=0.136$. 

\subsection{Identifiability}
In the previous two subsections, both inter-PUF and intra-PUF hamming distances and their mean values: $\mu_{inter}=3.703$ and $\mu_{intra}=0.136$ are calculated. Their comparison shows that $\mu_{inter}>\mu_{intra}$, which establishes that on average the PUF instances can be distinguished from each other. But the mean value does not depict the full story. Fig.~\ref{distribution}(b) shows the fitted distribution curves superimposed on each other. The brown curve (intra-PUF distribution) is skewed to the left and the blue curve (inter-PUF distribution) is skewed to the right and they mostly cover different regions. However, there is slight overlapping between them which is shown in the inset as a zoomed version of the overlapping area. Ideally, there should be no overlapping. But in a practical scenario, this overlapping region is the source of detection error.

From the definition of identifiability, a PUF class $\mathbb{P}$ is identifiable if it is reliable as well as unique, and if the probability of inter-PUF variation being greater than intra-PUF variation is
very high. Mathematically:
\begin{equation*}
    Probability(HD_{inter}>HD_{intra}) \approx 1
\end{equation*}

In previous two subsections, $94800\text{ } (=\frac{30\times80\times79}{2})$ intra-PUF distances and $435\text{ } (=\frac{30\times29}{2}$ inter-PUF distances have been calculated. Now, each of these inter-puf distances is compared with each of the intra-PUF distances that leads us to $435\times94800=41238000$ cases, among which,  $HD_{inter}>HD_{intra}$ is found in $41184206$ cases.  
\begin{equation*}
    \bm{Probability(HD_{inter}>HD_{intra}) = 0.9987}
\end{equation*}

This is a very high probability and close to $1$. This proves that RF-PUF has strong identifiability and this property along with reliability, uniqueness, constructability, and evaluability manifests RF-PUF as a distinct PUF class. This is the first-ever experimental validation of RF-PUF as a distinct and strong PUF class by itself.

\section{Possible Attack Models on RF-PUF}
\begin{figure}[t]
  \centering
  \includegraphics[width=0.8\linewidth]{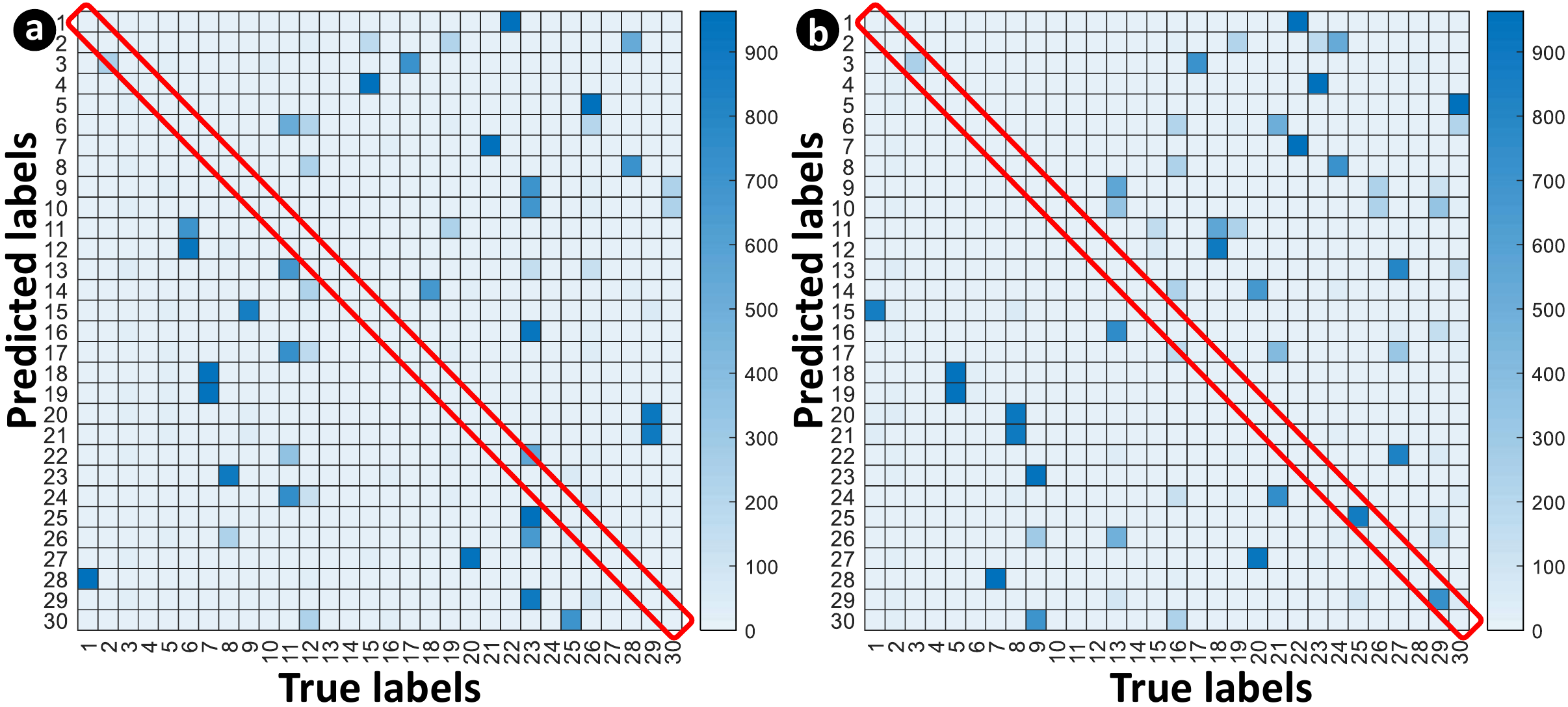}
  \caption{Heatmap of unsupervised learning in the attacker using k-means clustering. (a) The worst case accuracy of $0.09\%$ and (b) the best case accuracy of $6.8\%$. Repeated clustering for $1000$ times shows $3.63\%$ accuracy on average.}
  \label{heat}
\end{figure}

\begin{figure}[h]
  \centering
  \includegraphics[width=0.6\linewidth]{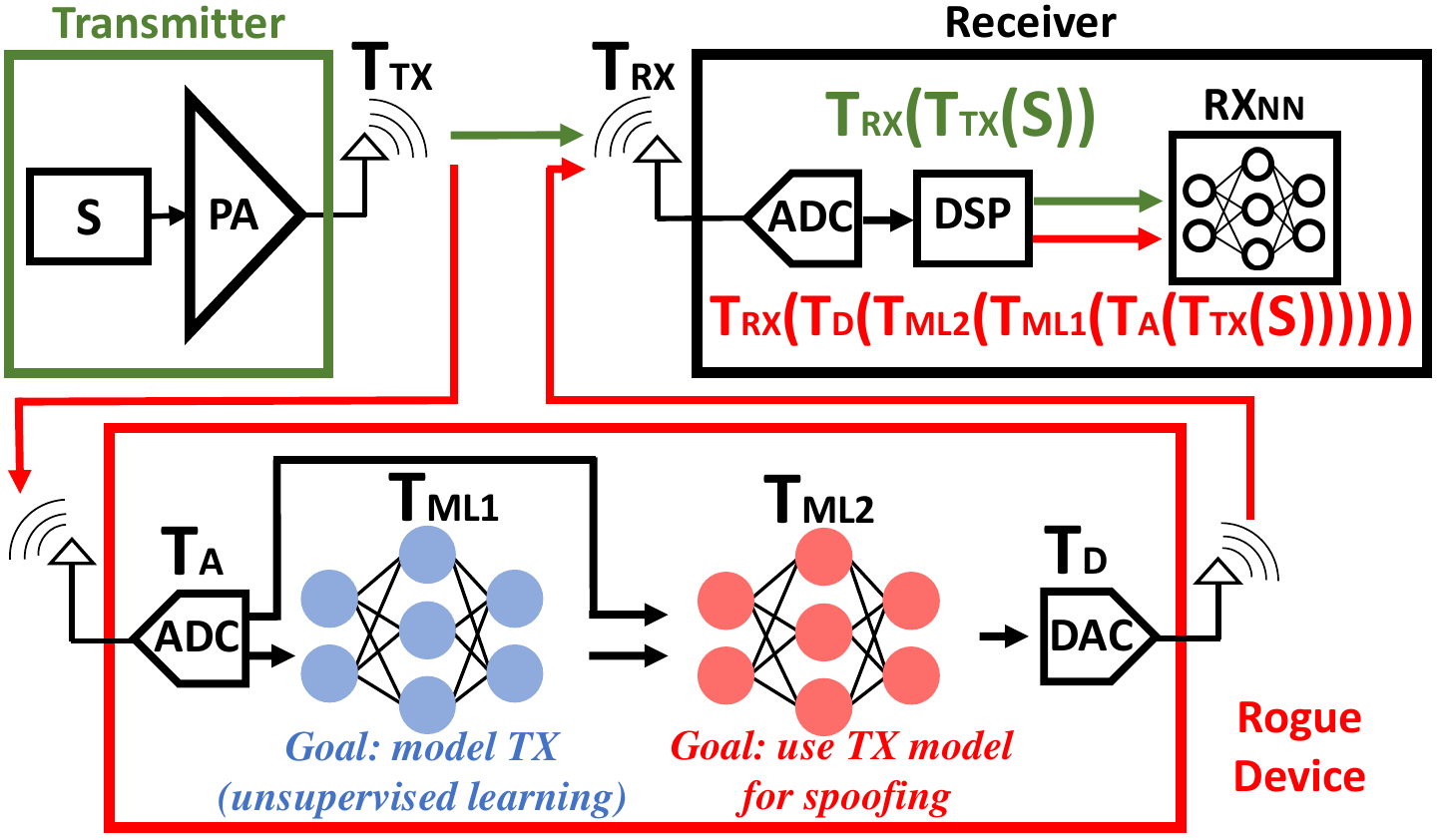}
  \caption{ML attack model. Adversary ML network cannot undo the effect of ADC and DAC which requires extremely high precision and resolution, making the attack impractical.}
  \label{ml_attack}
\end{figure}

RF-PUF does not store any digital key and hence, is not susceptible to malicious PUF models which assume that the adversary can have access to all the challenge-response pairs through a built-in logger software/implanted Trojan. However, there is a possibility of a machine learning-based attack that needs to be discussed (Fig.\ref{ml_attack}). For RF-PUF, ML attack is a two-step process:
\begin{itemize}
  \item \textbf{Step 1:} model/profile the victim TX (Unsupervised)
  \item \textbf{Step 2:} use that model for spoofing/replay attacks
\end{itemize}
In step 1, the rogue device tries to learn the feature/parameter values of the victim TX. Unlike the intended RX, this is an unsupervised problem for the attacker. We have utilized k-means clustering to divide the feature map into $30$ clusters and compare the predicted and true labels (Fig.~\ref{heat}). The process was repeated $1000$ times as k-means isn't unique without specific conditions. Our analysis shows that clustering achieves $\sim3.63\%$ accuracy on average, which is very close to the probability of random detection $(\frac{1}{30}=3.3\%)$. So, practically it is almost impossible to get the right feature value and label.

If somehow the attacker succeeds in step 1, then in step 2, the attacker needs to produce an RF signal that contains the same imperfections as the victim TX with high accuracy. This requires a high speed and high-resolution circuitry. Fig.~\ref{ml_attack} shows that the physical signature of the transmitter, $S$, goes through transformation $T_{TX}$ at TX and $T_{RX}$ at RX. The transformations in the attacker are $T_{A}$, $T_{ML1}$, $T_{ML2}$, and $T_{D}$ respectively. Full transformation for the original device is $T_{RX}$($T_{TX}$($S$)) and for the adversary is $T_{RX}$($T_{D}$($T_{ML2}$($T_{ML1}$($T_{A}$($T_{TX}$($S$)))))). The adversary ML2 framework needs to make these two transformations equal by undoing the effect of its ADC/DAC which requires almost infinite resolution, rendering it practically impossible (typical ADC/DAC are 8/16-bit). This Resolution limitation in ADC/DAC and bandwidth limitation in filters and other RF components also prevent replay attack, which requires the attacker to convert the TX signal in the digital domain, incorporate malicious contents and then transform it back into the RF domain with very high precision. Further analysis of precision requirements for a practical attack will be included in future work. The robustness of RF-PUF against malicious PUF model, ML attack, and replay attack proves its strong candidacy of employment for RF security.

\section{Conclusion}\label{conclude}
In this work, data collected from off-the-shelf commodity components ($30$ Xbee modules) were used to develop a new feature called the coefficient of frequency offset variation (COV) through PCA and moment analysis. The new feature leads to $95\%$ accuracy for a single hidden layer with $10$ neurons and $>99.8\%$ accuracy for a single hidden layer with $>50$ neurons, \textit{for the first time in literature without any assisting digital preamble}. The dataset containing 155.4 GB of data has also been released for public use. The design space has been explored and the effect of the wireless channel is analyzed to provide design insights. The scalability issue of simple ML algorithms for high accuracy has also been explored. The PUF properties of RF-PUF have been explored in detail. The inter-PUF and intra-PUF hamming distances are calculated and with their distribution, it has been shown that they have trivial overlapping. A detailed analysis reveals that the probability of $HD_{inter}>HD_{intra}$ = $0.9987$, which resonates with the claim that RF-PUF has a very high device authentication probability. Finally, possible important attack models are discussed and the robustness of RF-PUF against them is analyzed. This work experimentally validates RF-PUF with high accuracy, which can contribute to a secure authentication system using inherent physical signatures without extra power, area, or computational overhead on the resource-constrained IoT transmitter side.

\section*{Funding}
This work was supported in part by the National Science Foundation, under Grant 1935573.

\bibliographystyle{plainnat}

\begin{thebibliography}{00}
\bibitem{norton}
S. Symanovich, “The future of IOT: 10 predictions about the internet of things,” Norton, 28-Aug-2019. [Online]. Available: https://us.norton.com/internetsecurity-iot-5-predictions-for-the-future-of-iot.html. [Accessed: 14-Jan-2022]. 
\bibitem{DPA}
P. Kocher, J. Jaffe, and B. Jun, ``Differential power analysis," In Annual international cryptology conference (pp. 388-397), Springer, Berlin, Heidelberg, Aug., 1999.
\bibitem{EM_SCA}
J. J. Quisquater and D. Samyde, ``Electromagnetic analysis (ema): Measures and counter-measures for smart cards," In International Conference on Research in Smart Cards (pp. 200-210), Springer, Berlin, Heidelberg, Sep. 2001.
\bibitem{ML_SCA}
G. Hospodar, B. Gierlichs, E. D. Mulder, I. Verbauwhede, and J. Vandewalle, ``Machine learning in side-channel analysis: a first study," Journal of Cryptographic Engineering, 1(4), 293, 2001.
\bibitem{MFA_pat}
D. M. Ting, O. Hussain, and G. LaRoche, ``Systems and methods for multi-factor authentication," U.S. Patent No. 9,118,656, Washington, DC: U.S. Patent and Trademark Office, 2015.
\bibitem{MFA_survey}
A. Ometov, S. Bezzateev, N. Mäkitalo, S. Andreev, T. Mikkonen, and Y. Koucheryavy, ``Multi-factor authentication: A survey," Cryptography, 2(1), 1, 2018.
\bibitem{oauth}
“OAuth 2.0,” OAuth. [Online]. Available: https://oauth.net/2/. [Accessed: 14-Jan-2022].  
\bibitem{csrf1}
M. S. Siddiqui and D. Verma, ``Cross site request forgery: A common web application weakness," In 2011 IEEE 3rd International Conference on Communication Software and Networks, pp. 538-543, IEEE, May 2011.
\bibitem{csrf2}
A. Barth, C. Jackson, and J. C. Mitchell, ``Robust defenses for cross-site request forgery," In Proceedings of the 15th ACM conference on Computer and communications security, pp. 75-88, Oct., 2008.
\bibitem{RFPUF}
B. Chatterjee, D. Das, S. Maity and S. Sen, ``RF-PUF: Enhancing IoT Security Through Authentication of Wireless Nodes Using In-Situ Machine Learning,'' IEEE Internet of Things Journal, 2019.
\bibitem{PREAMBLE_OBFUS}
J. Chacko et al., ``Physical gate based preamble obfuscation for securing wireless communication,'' International Conference on Computing, Networking and Communications (ICNC), pp. 293-297, 2017.
\bibitem{RFfingerBT}
J. Hall, M. Barbeau, and E. Kranakis, ``Detecting Rogue Devices in Bluetooth Networks Using Radio Frequency Fingerprinting,'' International Conference on Communication and Computer Networks (CNN), 2006.
\bibitem{RFfingerFC}
P. Scanlon, I. O. Kennedy, and Y. Liu, ``Feature extraction approaches to RF Fingerprinting for Device Identification in Femtocells,'' Bell Labs Technical Journal, 2010.
\bibitem{RFfingerXbee}
T. J. Bihl, K. W. Bauer, and M. A. Temple, “Feature selection for RF fingerprinting with multiple discriminant analysis and using ZigBee device emissions,” IEEE Transactions on Information Forensics and Security, 2016.
\bibitem{RFfinger_4}
K. B. Rasmussen and S. Capkun, ``Implications of radio fingerprinting on the security of sensor networks," In 2007 Third International Conference on Security and Privacy in Communications Networks and the Workshops-Secure Comm 2007, pp. 331-340, IEEE, Sep., 2007.
\bibitem{RFfinger_5}
V. Brik, S. Banerjee, M. Gruteser, and S. Oh, ``Wireless device identification with radiometric signatures," in Proceedings of the 14th ACM International Conference on Mobile Computing and Networking, pp. 116-127, Sep., 2008.
\bibitem{RFfinger_6}
N. T. Nguyen, G. Zheng, Z. Han, and R. Zheng, ``Device fingerprinting to enhance wireless security using nonparametric Bayesian method," in 2011 Proceedings IEEE INFOCOM, pp. 1404-1412, Apr., 2011.
\bibitem{RFfinger_7}
T. D. Vo-Huu, T. D. Vo-Huu, and G. Noubir, ``Fingerprinting Wi-Fi devices using software defined radios," in Proceedings of the 9th ACM Conference on Security \& Privacy in Wireless and Mobile Networks, pp. 3-14, Jul., 2016.
\bibitem{RFfinger_8}
L. Peng, A. Hu, J. Zhang, Y. Jiang, J. Yu, and Y. Yan, ``Design of a hybrid RF fingerprint extraction and device classification scheme," IEEE Internet of Things Journal, 6(1), 349-360, 2018.
\bibitem{RFfinger_9}
F. Xie, H. Wen, Y. Li, S. Chen, L. Hu, Y. Chen, and H. Song, ``Optimized coherent integration-based radio frequency fingerprinting in Internet of Things," IEEE Internet of Things Journal, 5(5), 3967-3977, 2018.
\bibitem{RFfingerMACLayer}
Q. Xu, R. Zheng, W. Saad, and Z. Han, ``Device Fingerprinting in Wireless Networks: Challenges and Opportunities," in IEEE Communications Surveys \& Tutorials, vol. 18, no. 1, pp. 94-104, First quarter, 2016, doi: 10.1109/COMST.2015.2476338.
\bibitem{IMEI}
K. Kumar, P. Kaur, and G. N. D. U. Amritsar, ``Vulnerability detection of international mobile equipment identity number of smartphone and automated reporting of changed IMEI number," International Journal of Computer Science and Mobile Computing, 4(5), 527-533, 2015.
\bibitem{IP_hack}
Y. Wang and J. Yang, ``Ethical hacking and network defense: choose your best network vulnerability scanning tool," in 2017 31st International Conference on Advanced Information Networking and Applications Workshops (WAINA), pp. 110-113, Mar., 2017.
\bibitem{http_hack}
T. Chomsiri, ``HTTPS hacking protection," in 21st International Conference on Advanced Information Networking and Applications Workshops (AINAW'07), Vol. 1, pp. 590-594, May, 2007.
\bibitem{MAC_spoof}
B. Alotaibi and K. Elleithy, ``A new mac address spoofing detection technique based on random forests," Sensors, 16(3), 281, 2016.
\bibitem{PA_nonlinear}
S. S. Hanna and D. Cabric, ``Deep Learning Based Transmitter Identification using Power Amplifier Nonlinearity," 2019 International Conference on Computing, Networking and Communications (ICNC), Honolulu, HI, USA, 2019, pp. 674-680, doi: 10.1109/ICCNC.2019.8685569.
\bibitem{DL_1}
T. O’Shea and J. Hoydis, ``An introduction to deep learning for the physical layer," IEEE Transactions on Cognitive Communications and Networking, 3(4), 563-575, 2017.
\bibitem{DL_3}
T. Wang, C. K. Wen, H. Wang, F. Gao, T. Jiang, and S. Jin, ``Deep learning for wireless physical layer: Opportunities and challenges," China Communications, 14(11), 92-111, 2017.
\bibitem{DL_4}
C. Zhang, P. Patras, and H. Haddadi, ``Deep learning in mobile and wireless networking: A survey," IEEE Communications Surveys \& Tutorials, 21(3), 2224-2287, 2019.
\bibitem{DL_5}
Wang, X., Wang, X., \& Mao, S. (2018). RF sensing in the Internet of Things: A general deep learning framework. IEEE Communications Magazine, 56(9), 62-67.
\bibitem{IMS}
M. F. Bari, B. Chatterjee, K. Sivanesan, L. L. Yang, and S. Sen, ``High Accuracy RF-PUF for EM Security through Physical Feature Assistance using Public Wi-Fi Dataset," in 2021 IEEE MTT-S International Microwave Symposium (IMS) (pp. 108-111), Jun., 2021.
\bibitem{IEEE_802}
V. Brik, S. Banerjee, M. Gruteser, and S. Oh, ``Wireless device identification with radiometric signatures," in Proceedings of the 14th ACM International Conference on Mobile Computing and Networking (MobiCom ’08). Association for Computing Machinery, New York, NY, USA, 116–127, 2008. DOI:https://doi.org/10.1145/1409944.1409959
\bibitem{RF_PUF_Dataset}
M. F. Bari and S. Sen, “Sparclab RF-PUF Dataset,” GitHub. [Online]. Available: https://github.com/SparcLab/Sparclab-RF-PUF-Dataset. [Accessed: 14-Jan-2022].
\end{thebibliography}

\end{document}